\documentclass[useAMS,usenatbib]{mn2e}

\usepackage{graphicx}

\newcommand       \be           {\begin{equation}}
\newcommand       \ee           {\end{equation}}
\newcommand       \bea          {\begin{eqnarray}}
\newcommand       \eea          {\end{eqnarray}}
\newcommand       \apj          {ApJ}
\newcommand       \apjl         {ApJL}
\newcommand       \aap          {A\&A}
\newcommand       \nat          {Nature}
\newcommand       \mnras        {MNRAS}
\def\simlt{\mathrel{\hbox{\rlap{\hbox{\lower4pt\hbox{$\sim$}}}\hbox{$<$}}}}
\def\simgt{\mathrel{\hbox{\rlap{\hbox{\lower4pt\hbox{$\sim$}}}\hbox{$>$}}}}
\def\lesssim{\mathrel{\hbox{\rlap{\hbox{\lower4pt\hbox{$\sim$}}}\hbox{$<$}}}}

\def\gtrsim{\mathrel{\hbox{\rlap{\hbox{\lower4pt\hbox{$\sim$}}}\hbox{$>$}}}}

\title[]{The effects of $r$-process heating on fall-back accretion in compact object mergers}\author[B.~D. Metzger, A.~Arcones, E.~Quataert, G.~Mart\'{\i}nez-Pinedo]{B.~D. Metzger$^{1}$\thanks{E-mail:
bmetzger@astro.berkeley.edu}, A.~Arcones$^{2,3}$, E.~Quataert$^{1}$, G.~Mart\'{\i}nez-Pinedo$^{3}$ \\
$^{1}$Astronomy Department and Theoretical Astrophysics Center,
University of California, Berkeley, 601 Campbell Hall, Berkeley CA,
94720\\
$^{2}$Institut f\"{u}r Kernphysik, TU~Darmstadt,
Schlossgartenstr.~9, D-64289 Darmstadt, Germany\\
$^{3}$GSI Helmholtzzentrum f\"{u}r Schwerionenforschung,
Planckstr.~1, D-64291 Darmstadt, Germany\\
}

\begin{document}
\date{Accepted . Received ; in original form }
\pagerange{\pageref{firstpage}--\pageref{lastpage}} \pubyear{????}
\maketitle
\label{firstpage}

\begin{abstract}

We explore the effects of $r$-process nucleosynthesis on fall-back accretion in neutron star(NS)-NS and black hole-NS mergers, and the resulting implications for short-duration gamma-ray bursts (GRBs).  Though dynamically important, the energy released during the $r$-process is not yet taken into account in merger simulations.  We use a nuclear reaction network to calculate the heating (due to $\beta$-decays and nuclear fission) experienced by material on the marginally-bound orbits nominally responsible for late-time fall-back.  Since matter with longer orbital periods $t_{\rm orb}$ experiences lower densities, for longer periods of time, the total $r$-process heating rises rapidly with $t_{\rm orb}$, such that material with $t_{\rm orb} \gtrsim 1$ seconds can become completely unbound.  Thus, $r$-process heating fundamentally changes the canonical prediction of an uninterrupted power-law decline in the fall-back rate $\dot{M}_{\rm fb}$ at late times.  When the timescale for $r$-process to complete is $\gtrsim 1$ second, the heating produces a complete cut-off in fall-back accretion after $\sim 1$ second; if robust, this would imply that fall-back accretion cannot explain the late-time X-ray flaring observed following some short GRBs.  However, for a narrow, but physically plausible, range of parameters, fall-back accretion can {\it resume} after $\sim 10$ seconds, despite having been strongly suppressed for $\sim 1-10$ seconds after the merger.  This suggests the intriguing possibility that the gap observed between the prompt and extended emission in short GRBs is a manifestation of $r$-process heating.  

\end{abstract}


\section{Introduction}
\vspace{-0.1cm}

One of the most important discoveries made with the $Swift$ satellite is that short- and long-duration gamma-ray bursts (GRBs) originate from distinct stellar progenitors.  While long duration GRBs track ongoing star formation (e.g. Fructer et al.~2006) and result from the deaths of massive stars (e.g. Woosley $\&$ Bloom 2006), short-duration GRBs have now been localized to both early (Bloom et al.~2006; Berger et al.~2005) and late-type (Fox et al.~2005; Barthelmy et al.~2005) host galaxies, indicating a more evolved progenitor population (e.g. Prochaska et al.~2006).  

Although the origin of short GRBs remains unknown, the most popular
and well-studied central engine model is the binary merger of two
neutron stars (NS-NS; Eichler et al.~1989; Meszaros $\&$ Rees 1992;
Narayan et al.~1992) or a NS and a black hole (NS-BH; Paczy{\'n}ski
1991; Mochkovitch et al.~1993).  This model is qualitatively
consistent with both the host galaxy properties of short GRBs (Nakar
et al.~2006) and the notable lack of a bright associated supernova in
some well-studied cases (e.g. Hjorth et al.~2005).  Depending on the
detailed properties of the binary and the (unknown) supra-nuclear
density equation of state, NS-NS/NS-BH mergers result in a central
compact object (either a BH or hyper-massive NS) surrounded by a
compact $\sim 10^{-3}-0.1M_{\sun}$ disk (e.g. Ruffert et al.~1996;
Rosswog et al.~1999; Lee $\&$ Klu{\'z}niak 1999; Rosswog 2005; Faber
et al.~2006; Shibata $\&$ Taniguchi 2006; see Lee $\&$ Ramirez-Ruiz
2007 and Faber et al.~2009 for recent reviews).  The similarity
between the estimated accretion timescale of this remnant torus and
the typical duration of short GRBs ($T_{90} \sim 0.1-1$ seconds) has
also been used as evidence in favor of compact object merger models
(Narayan et al.~1992).

This clean picture has grown complex with the discovery that short GRBs are often followed by a ``tail'' of emission (usually soft X-rays) starting $\sim 10$ seconds after the GRB and lasting for $\sim 30-100$ seconds (Norris $\&$ Bonnell 2006; Gehrels et al.~2006).  While only $\sim 1/4$ of $Swift$ short bursts show evidence for this extended emission, the observational limits are consistent with most bursts possessing an X-ray tail with a fluence comparable to that of the prompt GRB (Perley et al.~2009).  Due to its rapid variability and other similarities to the prompt gamma-ray emission, the extended emission probably results from ongoing central engine activity.  However, such a significant energy output on long timescales is difficult to explain in NS-NS/NS-BH merger models, most notably because the accretion disk is blown apart by a powerful outflow after only a few seconds of viscous evolution (Metzger, Piro, $\&$ Quataert 2008, 2009; Beloborodov 2008; Lee et al.~2009).

One idea proposed for producing late-time activity is the
``fall-back'' of material ejected during the merger into highly
eccentric (but gravitationally-bound) orbits (Rosswog 2007; Lee $\&$
Ramirez-Ruiz 2007).  If the ejected mass $M$ has a distribution of
energies $dM/d|E| \propto |E|^{-\alpha}$, matter with $E < 0$ produces
a late-time accretion rate $\dot{M}$ onto the central object that
decreases as a power-law in time: $\dot{M} \propto
(dM/dE)(d|E|/dt_{\rm orb}) \propto t^{-[(5+2\alpha)/3]}$ (Rees 1988),
where $t_{\rm orb}$ is the bound ejecta's orbital period or fall-back
time.  Indeed, a number of NS-NS/NS-BH merger calculations find that
the fall-back at late-times may be sufficient to explain the extended
emission from short GRBs via continued accretion (Faber et al.~2006;
Rosswog 2007; Lee et al.~2009; see, however, Rossi $\&$ Begelman
2009).

Fall-back accretion models rely on the assumption that matter ejected onto orbits with fall-back times $t_{\rm orb}\gg 1$ second remains bound.  However, the gravitational binding energy of such long-period orbits is only
\be
|E| = \frac{GMm_{\rm n}}{2a} \simeq 1.0\left(\frac{M}{3M_{\sun}}\right)^{2/3}\left(\frac{t_{\rm orb}}{1{\,\rm s}}\right)^{-2/3}{\,\rm\frac{MeV}{nucleon}},
\label{eq:ebind}
\ee  
where $t_{\rm orb} = 2\pi(a^{3}/GM)^{1/2}$, $a$ is the semi-major axis of the orbit, and $M$ and $m_{\rm n}$ are the central object mass and nucleon mass, respectively.

The $unbound$ ejecta from NS-NS/NS-BH mergers has long been considered a promising source for producing very heavy elements via rapid neutron capture ($r$-process) nucleosynthesis (Lattimer $\&$ Schramm 1974, 1976; Eichler et al.~1989; Freiburghaus et al.~1999).  The total nuclear energy available via the $r$-process ($\sim 1-3$ MeV nucleon$^{-1}$; see eq.~[\ref{eq:enuc}]) greatly exceeds $|E|$ for orbits with $t_{\rm orb} \gtrsim 0.3-1$ second.  As a result, $r$-process heating could have a crucial impact on the properties of late-time fall-back, an effect that has not yet been taken into account.

In this paper, we examine the effects of $r$-process nucleosynthesis
on fall-back accretion in NS-NS/NS-BH mergers.  In
$\S\ref{sec:rprocessheating}$ we describe the nucleosynthesis that
occurs during the decompression from nuclear densities, and the extent
to which NS-NS/NS-BH merger simulations properly capture the resulting
energy release.  This motivates $\S\ref{sec:calculations}$, in which
we present calculations of $r$-process heating along orbits that are
nominally responsible for late-time fall-back.  In
$\S\ref{sec:discussion}$ we discuss our results and their
implications.

\vspace{-0.7cm}
\section{Decompression $\&$ $r$-Process Heating}
\label{sec:rprocessheating}
\vspace{-0.1cm}

Most of the material ejected when a NS is tidally disrupted originates
from the NS's neutron-rich outer core, which has a typical electron
fraction $Y_{e} \sim 0.1$ set by $\beta-$equilibrium under highly
degenerate conditions (Pethick $\&$ Ravenhall 1995; Haensel $\&$
Zdunik 1990a,b).  Since the temperature remains fairly low as the
ejecta expands (due to adiabatic losses), $Y_{e}$ probably remains low
($\sim 0.03-0.20$) during the decompression from nuclear densities
(Ruffert et al.~1997; Rosswog~2005).

Schematically, the nucleosynthesis of decompressing neutron-rich matter can be divided into two stages:  

(1) {\it Initial Decompression and Seed Formation} (density $\rho
\gtrsim \rho_{\rm drip} \sim 4\times 10^{11}$ g cm$^{-3}$).  During
the earliest phases of decompression, very neutron-rich nuclei form,
which rapidly emit neutrons as the material expands to lower density
(Lattimer et al.~1977; Meyer 1989).  Heavy ``seed'' nuclei are then
formed through $(n,\gamma)$ reactions and, possibly, through
charged-particle reactions in full nuclear statistical equilibrium
(NSE).  For example, Meyer (1989) finds seed nuclei with average
charges and masses $\bar{Z} \sim 40-70$ and $\bar{A} \sim 90-110$
(depending primarily on $Y_{e}$ and the expansion rate; cf.~Goriely et al.~2004), while
Freiburghaus et al.~(1999) finds seeds with $Z \approx 31-37$
and $A \approx 92-112$.  Since the seed nuclei and neutron mass
fractions are given by $X_{\rm s} = \bar{A}Y_e/\bar{Z}$ and $X_{\rm n}
= 1 - X_{\rm s}$,
respectively, the neutron mass fraction after initial decompression is large: $X_{\rm n} \sim 0.3-0.9$ for plausible ranges in the values of $\bar{A}$, $\bar{Z}$, and the electron fraction ($Y_{e} \sim 0.03-0.2$; Ruffert et al.~1997; Rosswog~2005). 

(2) {\it Rapid Neutron Capture ($r$-process)} ($\rho \lesssim \rho_{\rm drip}$).  Once the density decreases below neutron-drip, $\beta-$decay channels begin opening in full, and a conventional $r$-process begins (see, e.g.,  Cowan, Thielemann, $\&$ Truran 1991 and Meyer 1994 for reviews).  In the $r$-process, very heavy nuclei (with peaks at $A \sim 130$ and $200$) are formed when the seed nuclei rapidly capture the free neutrons remaining from stage 1.  This establishes an (n,$\gamma$) equilibrium, with $\beta-$decays driving the nucleosynthetic ``flow'' to larger $Z$ on longer timescales, with possible ``fission cycling'' between nuclei with $A \sim 280$ and $A \sim 130-140$. 

If NSE is assumed, the nuclear energy released when seeds form (stage 1) can be captured in numerical simulations of NS-NS/NS-BH mergers by employing an appropriate equation of state (EOS).  For instance, the NS-NS merger simulations of Rosswog et al.~(1999) use a Lattimer-Swesty (1991) EOS, which accounts for the possible presence of protons, neutrons, $\alpha-$particles, and a single average ``heavy'' nucleus.  As a result, they find that the ejected tidal tails ``explode'' due to the energy released as seed nuclei form.  Since the Shen EOS (Shen et al.~1998a,b) employed by Rosswog $\&$ Davies (2002; cf. Rosswog 2005) captures similar physics, the effects of seed nuclei formation are already taken into account in the fall-back estimates of Rosswog (2007).

However, the subsequent $r$-process has the potential to generate a comparable or greater amount of energy.  In particular, once all of the synthesized nuclei decay back to stable isotopes, the total nuclear energy released and available to heat the ejecta is
\be 
\Delta E_{\rm r} \simeq (1-f_{\nu})\left[\left(\frac{B}{A}\right)_{r} - X_{\rm s}\left(\frac{B}{A}\right)_{s} - X_{\rm n}\Delta_{\rm n}\right],
\label{eq:enuc}
\ee
where $\Delta_{\rm n} = (m_{\rm n}-m_{\rm p})c^{2} =$ 1.293 MeV is the neutron-proton mass difference, and $\left(\frac{B}{A}\right)_{s,r}$ are appropriately-averaged binding energies for the seed and $r$-process nuclei, respectively.
The factor $f_{\nu}$ is the fraction of the nuclear energy lost to neutrinos and is $\sim 0.5$ (see $\S\ref{sec:calcdetails}$).  Using typical values of $\left(\frac{B}{A}\right)_{r} \approx 8$ MeV nuc$^{-1}$ and $\left(\frac{B}{A}\right)_{s} \approx 8.7$ MeV nuc$^{-1}$, we estimate that $\Delta E_{\rm r} \approx 1-3$ MeV nuc$^{-1}$ for $X_{\rm n}$ in the range $\sim 0.3-0.9$. 

Comparing $\Delta E_{\rm r}$ with the binding energy of the fall-back
material (eq.~[\ref{eq:ebind}]), we conclude that if the $r$-process
goes to completion, it will strongly affect the $dynamics$ of orbits
with $t_{\rm orb} \gtrsim 0.3-1$ seconds.  Unlike seed nuclei
formation, the effects of $r$-process heating cannot be readily
incorporated into merger simulations, in part because most of the
energy is released on length and time scales exceeding that which can
be presently simulated.  More importantly, because the $r$-process is
a non-equilibrium process involving a large number of exotic nuclei, its
study requires a complex reaction network, which would be prohibitive
to include in multi-dimensional simulations.  In the next section we
explore the $r$-process heating of bound ejecta by performing
nucleosynthesis calculations along a few representative Lagrangian
density trajectories.

\vspace{-0.5cm}
\section{Nucleosythesis Calculations}
\label{sec:calculations}

\subsection{Density Trajectories}
\label{sec:trajectories}

When the less massive NS is tidally disrupted during a NS-NS or NS-BH merger, a portion of the stellar material is ejected into one or two long tidal tail through the outer Lagrange points (Lattimer $\&$ Schramm 1974).  This material is imparted with a distribution of energies (or, equivalently for bound material, semi-major axes $a$).  Initially, all of the ejecta (bound and unbound) is approximately spatially coincident (at an assumed pericenter distance $r_{p} \approx 10^{7}$ cm) and shares a common density during decompression.  Thus, during the early expansion we use the density trajectory $\rho(t)$ corresponding to the $unbound$ ejecta studied in $r$-process calculations by Freiburghaus et al.~(1999) and taken from the NS-NS merger simulations of Rosswog et al.~(1999).  

On later timescales, material with energy $E<0$ and fall-back time
$t_{\rm orb}\propto |E|^{-3/2}$ spatially decouples from the unbound
ejecta, once their orbits approach apocenter.  Motivated by
simulations (e.g. Rosswog 2007) and theoretical considerations (Rees
1988), we assume $dM/d|E| \propto$ constant, corresponding to a
fall-back rate of
\begin{eqnarray} 
  \dot{M_{\rm fb}} &\simeq& \left(\frac{dM}{dE}\right)\left(\frac{d|E|}{dt_{\rm orb}}\right) \nonumber \\
  &\approx& 10^{-2}M_{\sun}{\,\rm s^{-1}}\left(\frac{dM/dE}{10^{-2}M_{\sun}/{\rm MeV\,nuc^{-1}}}\right)\left(\frac{t_{\rm orb}}{1\,{\rm s}}\right)^{-5/3},
\label{eq:mdot1}
\end{eqnarray}
where we normalize $dM/dE$ so that $\sim 10^{-2} M_{\sun}$ returns to the central object on a timescale $\gtrsim 1$ second (absent the effects of $r$-process heating), as is required in models that attribute late-time X-ray tails from short GRBs to fall-back (e.g. Rosswog 2007; Faber et al.~2006; Lee et al.~2009). 

\begin{figure}
\resizebox{\hsize}{!}{\includegraphics[]{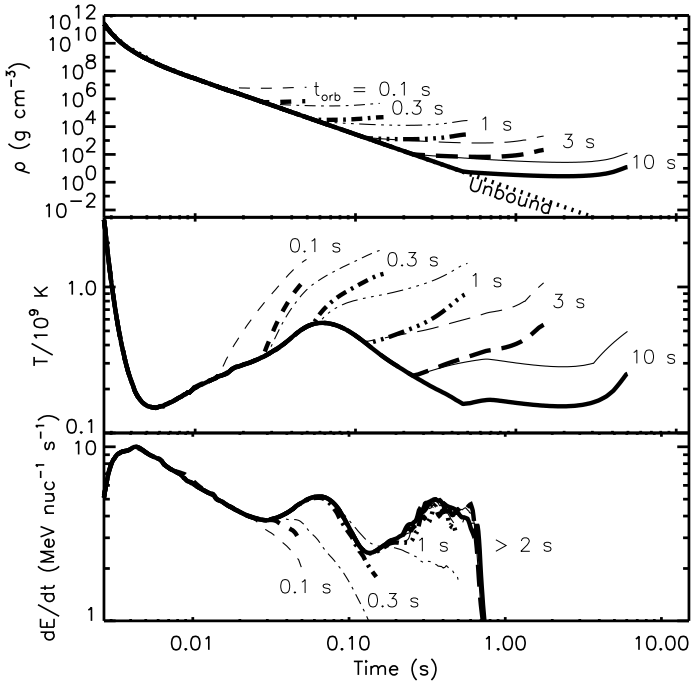}}
\caption{$r$-process nucleosynthesis in bound ejecta from compact
  object mergers.  ($Upper$ $Panel$) Lagrangian density trajectories
  $\rho(t)$ employed in our nucleosynthesis calculations, shown for
  ejecta with different initial orbital periods $t_{\rm orb} = $ 0.1 s ($short$ $dashed$ $lines$), 0.3 s ($dash-dot$ $lines$), 1 s ($triple-dot-dash$ $lines$), 3 s ($long$ $dashed$ $lines$), and 10 s ($solid$ $lines$) and for two values for the orbits' solid angle $\Delta\Omega = 4\pi, 4\pi/10$ ($darker$ $and$ $lighter$ $shaded$ $lines,$ $respectively$).  The trajectory of the unbound
  material (which all orbits share at early times) is shown with a
  dotted line.  ($Middle$ $Panel$) Temperature evolution for the
  trajectories shown in the upper panel.  ($Bottom$ $Panel$) Total
  $r$-process heating rate (due to $\beta-$decays and nuclear fission)
  for the trajectories shown in the upper panel, assuming that 1/2 of the energy is lost to neutrinos (see $\S\ref{sec:calcdetails}$).}
\label{fig:trajectories}
\end{figure}

The mass-flux along a series of trajectories with a given fall-back time $t_{\rm orb}$ can also be written as
\be
\dot{M} = \Delta\Omega \rho v r^{2},
\label{eq:mdot2}
\ee
where $r(t)$ and $v(t)$ are the radius and velocity of the orbit, and $\Delta\Omega(t)$ is the spread in solid angle of bound fluid elements.  By equating equations (\ref{eq:mdot1}) and (\ref{eq:mdot2}) we obtain the $late$-$time$ density trajectory.  Our ignorance of the details of the merger (which depends on uncertainties such as the NS EOS) and the effects of nuclear energy input on the ejecta trajectories are parameterized with $\Delta\Omega(t)$.  For simplicity we assume that $\Delta\Omega$ is constant with time because our results are relatively insensitive to this choice.  We choose a relatively large value for $\Delta\Omega \sim 4\pi/10 - 4\pi$, motivated by the large dispersion in the bound ejecta's orbital parameters expected to result from the explosive energy release during seed formation ($\S\ref{sec:rprocessheating}$) and the subsequent $r$-process heating.  We only calculate heating until the orbits reach apocenter because we are primarily interested in the total energy release, and $r$-process heating decreases rapidly once material re-compresses on its return to pericenter.  The top panel of Figure \ref{fig:trajectories} shows the density trajectories employed in our calculations for a variety of fall-back times.

\vspace{-0.5cm}
\subsection{Network Calculations}
\label{sec:calcdetails}

We use a dynamical $r$-process network calculation (Martinez-Pinedo
2008; Petermann et al.~2008) that includes neutron captures, photodissociations, $\beta-$decays, and fission reactions.  The latter includes
contributions from neutron induced fission, $\beta$ delayed fission,
and spontaneous fission.  All heating is self-consistently added to
the entropy of the fluid following the procedure of Freiburghaus et
al.~(1999). The change of temperature is determined using the Timmes
equation of state (Timmes \& Arnett 1999).  Although our calculation does not explicitly account for the energy loss from $\beta-$decays into escaping neutrinos, we take this into account by artificially decreasing the heating rate by a factor $1/2$.  This is justified because most of the heating results from $\beta-$decays and the energy released is shared approximately equally between electrons (which thermalize) and neutrinos.   

In addition to $\rho(t)$, the initial temperature $T$, electron
fraction $Y_{e}$, and seed nuclei properties ($\bar{A}$,$\bar{Z}$) are
specified for a given calculation.  We assume an initial temperature
$T = 4\times 10^{9}$ K, although the subsequent $r$-process heating is
not particularly sensitive to this choice (Meyer 1989; Freiburghaus et al.~1999).  We also assume $Y_{e} = 0.1$, $\bar{Z} \simeq 36$, $\bar{A} \simeq
118$ (see $\S\ref{sec:rprocessheating}$).  Varying the
electron fraction and the properties of the seed nuclei will
quantitatively affect the subsequent $r$-process heating (e.g. through the total available energy; eq.~[\ref{eq:enuc}]); the implications of this are discussed in $\S\ref{sec:implications}$.

\subsection{Results}
\label{sec:results}

\begin{figure}
\resizebox{\hsize}{!}{\includegraphics[]{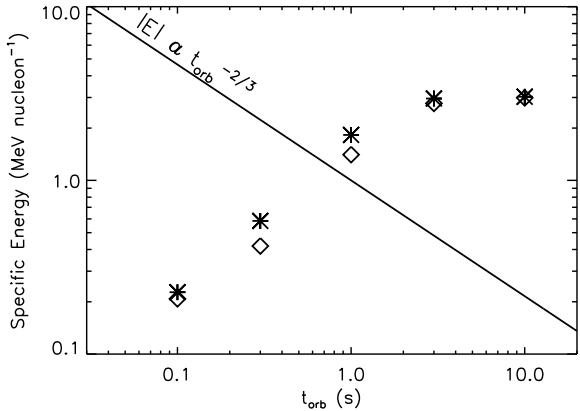}}
\caption{Total $r$-process heating $\Delta E$ as a function of initial orbital
  period (or nominal fall-back time) $t_{\rm orb}$ for the
  nucleosynthesis calculations shown in Figure
  $\ref{fig:trajectories}$, shown for $\Delta\Omega = 4\pi$
  ($asterisks$) and $4\pi/10$ ($diamonds$).  The binding energy of the
  orbit $|E|$ (eq.~[\ref{eq:ebind}]) is shown for comparison with a
  solid line (for an assumed $M = 3M_{\sun}$ central object).}
\label{fig:edot}
\end{figure}

The middle and bottom panels of Figure \ref{fig:trajectories} show the time evolution of the temperature $T$ and heating rate $\dot{E}$, respectively. The temperature initially decreases rapidly due to the adiabatic expansion.  However, as the expansion rate decreases, nuclear energy generation reheats the ejecta to $T \approx 5\times 10^{8}$ K.  At this point a difference develops between trajectories with short orbital periods ($t_{\rm orb} \lesssim 1$ s) and long orbital periods ($t_{\rm orb} \gtrsim 1$ s).  For short orbital periods the density approaches an approximately constant value relatively quickly.  At constant density all of the generated energy is used to increase the temperature, which reaches a value $T \gtrsim 10^{9}$ K.  At these high temperatures photodissociation reactions become important and the $r$-process path moves closer to stability.  This decreases the beta-decay rates and Q-values, and, consequently, the heating rate decreases. For the longest period orbits ($t_{\rm orb} \gtrsim 1$ s), there is a balance between energy generation and adiabatic losses, which keeps the temperature almost constant. 

Also note that $\dot{E}$ remains relatively constant in the range $\sim 2-5$ MeV nuc$^{-1}$ s$^{-1}$ throughout most of the sampled trajectories for $t \lesssim 1$ second.  This illustrates that the total $r$-process heating $\Delta E = \int_{0}^{t_{\rm orb}/2}\dot{E}dt$ is dominated by relatively late times in the orbit, and that $\Delta E$ is approximately proportional to $t_{\rm orb}$.  Note that at $t\approx 0.7$ second $\dot{E}$ sharply decreases once neutrons are exhausted and the $r$-process is effectively complete.

The total $r$-process heating is shown explicitly in Figure \ref{fig:edot}, which plots $\Delta E$ for the trajectories from Figure \ref{fig:trajectories} as a function of orbital period.  The orbital binding energy $|E| \propto
t_{\rm orb}^{-2/3}$ (eq.~[\ref{eq:ebind}]) is plotted for comparison
with a solid line.  Note that $\Delta E$ rises rapidly with $t_{\rm orb}$ for both values of $\Delta \Omega$, before saturating at $\Delta E_r \approx 3$ MeV (eq.~[\ref{eq:enuc}]) for $t_{\rm orb} \gtrsim 2$ seconds; material with $t_{\rm orb} \gtrsim 1$ second experiences sufficient heating to become unbound.

\vspace{-0.5cm}
\section{Discussion and Implications}
\label{sec:discussion}

\subsection{Fall-Back Accretion}
\label{sec:cutoff}

Figure \ref{fig:edot} illustrates that orbits with initial periods
exceeding a fraction of a second experience sufficient $r$-process
heating to become unbound.  However, this does not by itself guarantee
a suppression in the fall-back rate $\dot{M}_{\rm fb}$ at late times
because the marginally-bound material ejected by the $r$-process (with
initial orbital energy $|E_{\rm i}|\ < \Delta E$) could in principle
simply be replaced by material that was initially more tightly bound
($|E_{\rm i}| \gtrsim \Delta E$).  It is also important to understand
how the effects of $r$-process heating differ from that of seed
nucleus formation; both release comparable amounts of the energy, yet
seed formation is already incorporated in merger calculations and does
not produce a sharp cut-off in the fall-back rate (Rosswog 2007).

Although our $r$-process calculations assumed ballistic ejecta
(\S\ref{sec:trajectories}), this approximation is no longer valid once
significant energy is added to the orbit (i.e., if $\Delta E \sim
|E_{\rm i}|$).  Because the ejecta is optically-thick, most of the
deposited thermal energy is transferred into kinetic energy via
adiabatic expansion.  This puts the ejecta on less-bound orbits which,
due to their lower densities and longer periods, experience even more
$r$-process heating (Fig.~\ref{fig:trajectories}).  This suggests that
the $r$-process may lead to a run-away in which all material with an
initial orbital period exceeding a threshold value will acquire
sufficient energy to become unbound.

To explore these issues quantitatively, we consider a toy model to
calculate $\dot{M}_{\rm fb}$ including the effects of $r$-process
heating.  We consider an ensemble of mass elements $dM$ distributed
with initial energies $E_{i}$ according to $dM/d|E_{i}| \propto$
constant (as in eq.~[\ref{eq:mdot1}]).  The final energy of each mass
element at apocenter is given by \be E_{\rm f} = E_{i} +
\int_{0}^{t_{\rm orb}(E)/2}\dot{E}dt,
\label{eq:ef}
\ee where $\dot{E}(t)$ is the heating rate along the orbit.  As
discussed in $\S\ref{sec:results}$ and shown in Figures
\ref{fig:trajectories} and \ref{fig:edot}, $\dot{E}$ is roughly
constant in time along an orbit (for $t \lesssim 1$ s).  Thus, we make the simplifying
assumption that $\dot{E} = \Delta E_{\rm r}/t_{\rm heat}$ for $t \le
t_{\rm heat}$ and $\dot{E}=0$ for $t > t_{\rm heat}$, where $\Delta
E_{\rm r}$ is the total available $r$-process energy
(eq.~[\ref{eq:enuc}]) and $t_{\rm heat}$ is the timescale for
$r$-process heating (which we leave as a free parameter).  Our results presented below are relatively
insensitive to the precise functional form of $\dot{E}$ prior to
$t_{\rm heat}$ provided that the total heating is dominated by late
times in the orbit.

If the final energy at apocenter $E_{\rm f}$ is $<0$ for a given mass
element, it remains bound despite $r$-process heating, with a new
fall-back time which we approximate as $t_{\rm orb}(E_{\rm f})$.  If, on the other
hand, $E_{\rm f} > 0$ the particle is unbound from the central object
and does not contribute to late-time accretion.  Note that because the
upper limit of integration in equation ($\ref{eq:ef}$) increases with
the orbital energy, this model allows for the run-away effect
described above.

\begin{figure}
\resizebox{\hsize}{!}{\includegraphics[]{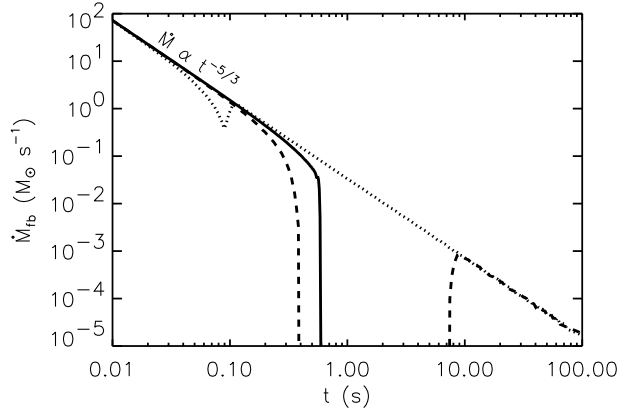}}
\caption{Fall-back rate $\dot{M}_{\rm fb}$ in NS-NS/NS-BH mergers
  including the effects of $r$-process heating, as calculated using
  the toy model described in $\S\ref{sec:cutoff}$.  Each model
  assumes a $M = 3 M_{\sun}$ central object and that the total
  available $r$-process energy is $\Delta E_{\rm r} = 3$ MeV.
  Different line styles correspond to different heating timescales:
  $t_{\rm heat} = 0.1$ seconds ($dotted$ $line$), 0.86 seconds ($dashed$
  $line$), and 3 seconds ($solid$ $line$).}
\label{fig:mdot}
\end{figure}

Figure $\ref{fig:mdot}$ shows our results for $\dot{M}_{\rm fb}$ with
$r$-process heating, calculated for fixed $\Delta E_{\rm r} = 3$ MeV
and for several values of $t_{\rm heat}$.  Because tightly-bound
material with short orbital periods experiences little $r$-process
heating, $\dot{M}_{\rm fb}$ at early times is unaffected by the
$r$-process and decreases at the canonical rate $\propto t^{-5/3}$ for
all values of $t_{\rm heat}$.  At late-times, however, there is a
bifurcation in the behavior of $\dot{M}_{\rm fb}$: short heating times
($t_{\rm heat} \ll 0.9$ s) lead to a relatively uninterrupted power law
decline, while long heating times ($t_{\rm heat} > 0.9$ s) produce a
sharp cut-off in the fall-back rate.

The origin of this bifurcation can be understood by noting the  
existence of a critical orbit distinguishing material that necessarily  
experiences the full heating available $\Delta E_{r}$ from those that may not.  When r-process heating is active, the energy of an orbit evolves as $E = -|E_{\rm i}| + \dot{E}t$.  As time increases, the magnitude of the orbital energy decreases, i.e., $|E|$ goes down, and thus the orbital period increases.  There is a critical orbit for which the $r$-process heating leads to the orbital period increasing so rapidly with time that the orbit can never actually reach apocenter ($t = t_{\rm orb}/2$) as long as $r$-process heating is active.   These orbits necessarily receive the full $r$-process heating $\Delta E_{r}$.   This critical orbital energy $E_{\rm c}$ can be determined by setting $dt/d|E| = 1/2(dt_{\rm orb}/d|E|)$ using equation (\ref{eq:ebind}) and $E = -|E_{\rm i}| + \dot{E}t$.   This implies

\be |E_{\rm c}| \simeq 1.35{\,\rm MeV}\left(\frac{M}{3M_{\sun}}\right)^{2/5}\left(\frac{\Delta E_{\rm r}}{3\,{\rm MeV}}\right)^{2/5}\left(\frac{t_{\rm heat}}{1\,{\rm s}}\right)^{-2/5}.
\label{eq:ec}
\ee 

The {\it initial} orbital energy for which $E_{\rm c}$ is just reached at  
apocenter is given by $E(t = t_{\rm orb}/2) = E_{\rm i} + \dot{E}t_{\rm orb}/2 = E_{\rm c}$, which implies $|E_{\rm i,c}| = 5|E_{\rm c}|/3$.  Note that equation (\ref{eq:ec}) can be estimated on dimensional grounds by solving for $E_{\rm i}$ such that $|E_{\rm i}| \approx (\Delta E_{r}/t_{\rm heat})t_{\rm orb}(E_{\rm i})/2$.

There is a bifurcation in the behavior for $|E_{\rm i}| < |E_{\rm i,c}|$ and $|E_{\rm i}| > |E_{\rm i,c}|$.  Orbits with $|E_{\rm i}| < |E_{\rm i,c}|$ necessarily receive the full heating and so have $E_{\rm f} = -|E_{\rm i}| + \Delta E_{\rm r}$, while those with $|E_{\rm i}| > |E_{\rm i,c}|$ may or may not (see below).  The orbital period (or fall-back time) corresponding to $|E_{\rm c}|$ is given by
\be t_{\rm orb,c} \simeq 0.6\,{\rm
  s}\left(\frac{M}{3M_{\sun}}\right)^{2/5}\left(\frac{\Delta E_{\rm
      r}}{3\,{\rm MeV}}\right)^{-3/5}\left(\frac{t_{\rm heat}}{1\,{\rm
      s}}\right)^{3/5}, \ee from which it follows that \be
\frac{t_{\rm heat}}{t_{\rm orb,c}} \simeq 1.7
\left(\frac{M}{3M_{\sun}}\right)^{-2/5}\left(\frac{\Delta E_{\rm
      r}}{3\,{\rm MeV}}\right)^{3/5}\left(\frac{t_{\rm heat}}{1\,{\rm
      s}}\right)^{2/5}.
\label{eq:ratio}
\ee

If $t_{\rm heat} \ll t_{\rm orb,c}$ then any material with $t_{\rm
  orb} > t_{\rm orb,c}$ has already experienced the full heating
$\Delta E_{\rm r}$ earlier in its orbit.  In this case, there is a
slight decrease in $\dot{M}_{\rm fb}$ around the time at which $t_{\rm
  orb} \sim t_{\rm heat}$, but there is no significant interruption in
the fall-back rate; this corresponds to $t_{\rm heat} = 0.1$ s in
Figure \ref{fig:mdot}.  In particular, $\dot{M}_{\rm fb}$ still
decreases as a power-low $\propto t^{-5/3}$ at late times because
adding a constant energy to each $dM$ simply renormalizes the energy
scale when $dM/dE$ is flat.  This explains why seed nucleus formation,
which occurs on roughly the initial expansion timescale $\sim$
milliseconds $\ll t_{\rm orb,c}$, has little effect on the rate that $\dot{M}_{\rm
  fb}$ decreases at late times.

On the other hand, if $t_{\rm heat} \gg t_{\rm orb,c}$ then an
absolute cut-off in $dM/d|E_{\rm f}|$ (and hence $\dot{M}_{\rm fb}$)
occurs for $E_{\rm f} \gtrsim E_{\rm c}$, corresponding to times $t
\gtrsim t_{\rm orb,c}$.  This case is well-illustrated by the $t_{\rm
  heat} = 3$ s model in Figure \ref{fig:mdot}.

In intermediate cases, when $t_{\rm heat} \sim t_{\rm orb,c}$ (i.e.,
$|E_{\rm i,c}| \sim \Delta E_r$), there is still cut-off in the
accretion, but material with $|E_{\rm i}| \gtrsim |E_{\rm i,c}|$ may
remain marginally-bound despite the extra energy it receives, thus
leading to a temporal gap in $\dot{M}_{\rm fb}(t)$.  This is
illustrated by the $t_{\rm heat} = 0.86$ s model in Figure
\ref{fig:mdot}, which shows a long delay between the cut-off in
accretion at $t \approx 0.4$ s and its resumption at $t \approx 10$ s.
We find, however, that the gap only exists for a fairly narrow range
of parameters, and, when present, its width $\Delta t_{\rm gap}$ is exponentially sensitive to $t_{\rm heat}/t_{\rm orb,c}$; e.g., increasing $t_{\rm heat}/t_{\rm orb,c}$ from 1.3 to 1.7 increases $\Delta t_{\rm gap}$ from $\sim 1$ to $\sim 100$ s.

From our calculations in $\S\ref{sec:calculations}$ we find that $\Delta E_{\rm r} \approx
3$ MeV and $t_{\rm heat} \approx 0.7$ s, which corresponds to $t_{\rm heat} \sim 1.5 t_{\rm orb,c}$.  This is at the boundary between the ``absolute cut-off'' and ``intermediate'' regimes described above and shown in Figure \ref{fig:mdot}.  We discuss the implications of this result in the next section.

\vspace{-0.51cm}
\subsection{Implications for the Origin of Short GRBs with Extended Emission}
\label{sec:implications}

Our primary conclusion is that fall-back accretion following
NS-NS/NS-BH mergers is suppressed on timescales exceeding $\sim 0.3-1$ seconds due to $r$-process heating.  This result has important
implications for the origin of extended X-ray emission observed $\sim
10-100$ seconds following some short GRBs.  As discussed in the
Introduction, standard merger models have difficulty explaining
activity on such a late timescale, which has lead to the suggestion
that late-time flaring is powered by fall-back (Faber et al.~2006;
Rosswog 2007; Lee et al.~2009).  If long heating timescales and/or high values of $\Delta E_{\rm r}$ obtain (such that $t_{\rm heat} > 2 t_{\rm orb,c}$), our results strongly disfavor this explanation because the cut-off in $\dot{M}_{\rm fb}$ is absolute: very little material returns to the central object at late times (see the solid line in Fig.~\ref{fig:mdot}).

This conclusion changes, however, if $\Delta E_{\rm r}$ is lower and/or $t_{\rm heat}$ is shorter, such that $t_{\rm heat} \lesssim 2 t_{\rm orb,c}$.  In this case, the $r$-process produces a temporal gap in the fallback rate instead of an absolute cut-off (Fig.~\ref{fig:mdot}).  Intriguingly, the extended emission following short GRBs shows a lull of $\sim 3-10$ seconds between the end of the GRB and the beginning of the extended emission (e.g. Norris $\&$ Bonnell 2006; Gehrels et al.~2006; Perley et al.~2009).  Attributing this delay to $r$-process heating appears, however, to require fine-tuned parameters: $t_{\rm heat}/t_{\rm orb,c}$ must be between $\sim 1.5-1.7$ in order for the temporal gap to have a duration of $3-30$ seconds.  Nevertheless, typical parameters for $r$-process heating are not far from the critical condition $t_{\rm heat} \sim t_{\rm orb,c}$.  

\begin{figure}
\resizebox{\hsize}{!}{\includegraphics[]{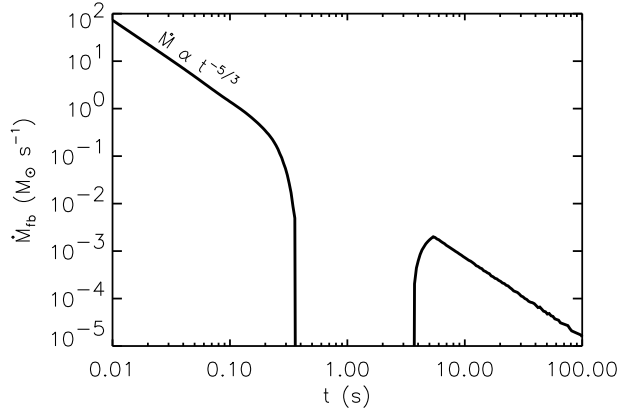}}
\caption{Fall-back rate $\dot{M}_{\rm fb}$ calculated using the model from $\S\ref{sec:cutoff}$ taking $\dot{E}(t)$ directly from our $r$-process calculations in Figure $\ref{fig:trajectories}$ for the $t_{\rm orb} = 10$ s orbit.}
\label{fig:mdot_calc}
\end{figure}

To illustrate this explicitly, Figure \ref{fig:mdot_calc} shows $\dot{M}_{\rm fb}(t)$ calculated using the simple model described in $\S\ref{sec:cutoff}$, but using $\dot{E}$ directly from our $r$-process calculations in Figure $\ref{fig:trajectories}$ for $t_{\rm orb} = 10$ seconds.   For these particular parameters, the exact $\dot{E}(t)$ suppresses fall-back accretion from $\sim 0.4-4$ seconds, but fall-back then resumes at late times.   Although the properties of the gap are quite sensitive to the parameters of the r-process heating, the presence of a gap may in fact be rather robust, because the parameters on which $t_{\rm heat}$ and $t_{\rm orb,c}$ depend ($\Delta E_{\rm r}, M, t_{\rm heat}$) may not vary substantially from event to event.  For instance, the outer portions of a NS ejected during a NS-NS/NS-BH merger are probably highly neutron-rich and likely remain so during the expansion (Ruffert et al.~1997).  For low $Y_{e}$ the total energy released by the $r$-process asymptotes to the value $\Delta E_{\rm r} \approx 0.5[(B/A)_{r} - \Delta_{\rm n}] \approx 3.3$ MeV (provided that $Y_{e}$ is not sufficiently low that the $r$-process freezes-out before all neutrons are captured).  Furthermore, for a merger to produce a significant accretion disk (as is required to produce a GRB), the mass of the central black hole or NS probably must lie in the relatively narrow range $M \sim 3-10M_{\sun}$ (e.g. Rantsiou et al.~2009).  Finally, a heating timescale of $\approx 1$ second also appears to be robust, relatively independent of uncertainties such as $\Delta \Omega$ (see Fig.~\ref{fig:edot}); $t_{\rm heat} \approx 1$ second was also found in an independent $r$-process calculation employing somewhat different assumptions and physics (see Goriely et al.~2004; their Fig.~3).  Further work is thus clearly required to assess the possibility that the gap between the prompt and extended emission in short GRBs is a manifestation of $r$-process heating.  In particular, given the sensitivity of the late-time fall-back rate to the $r$-process heating, it is important to consider a range of initial $Y_{e}$ and seed nuclei in the $r$-process calculations, and to improve on the simple model in $\S\ref{sec:cutoff}$ by carrying out hydrodynamic calculations of fall-back using simplified models of the $r$-process heating rate.

If future work indicates that $r$-process heating produces a robust sharp cut-off in $\dot{M}_{\rm fb}$ (with no late-time resumption), our results may instead suggest that the central engine in some short GRBs is a long-lived
NS rather than a BH, because the former could remain active even in
the absence of surrounding matter.  For instance, the massive central
object that forms in a NS-NS merger could be temporarily supported by
differential rotation (e.g., Baumgarte et al.~2000; Duez et al.~2004,
2006) or remain stable indefinitely if it loses sufficient mass via a
centrifugally-driven outflow (e.g. Thompson et al.~2004; Dessart et
al.~2008).  Another possibility for producing a stable NS is via the
accretion-induced collapse (AIC) of a white dwarf.  In either AIC or NS-NS mergers, if the rapidly-rotating NS is strongly
magnetized, its electromagnetic spin-down could plausibly power the
observed extended emission (Usov 1992; Metzger, Quataert, $\&$
Thompson 2008).  If AIC occurs following a double white dwarf merger,
late-time emission could also be powered by the accretion of material
left over after the merger, which has an accretion timescale $\sim
100$ seconds (Metzger, Quataert, $\&$ Thompson 2008).

\vspace{-0.5cm}
\section*{Acknowledgments}
{
We thank R.~Hix, J.~Lattimer, W.~Lee, and T.~Rauscher for helpful
discussions and information.  BDM and EQ were partially supported by the Packard Foundation.  AA and GMP were partially supported by the Deutsche Forschungsgemeinschaft through contract SFB 634 and by the Helmholtz Alliance of the Extreme Matter Institute (EMMI).
}

\label{lastpage}

\end{document}